\documentclass[aps,prb,preprint,showpacs,amsmath,amssymb]{revtex4-1}
\usepackage{graphicx}
\usepackage{dcolumn}
\usepackage{bm}
\usepackage{footnote}

\newcommand{\BTO}{$\text{BaTiO}_3$}
\newcommand{\STO}{$\text{SrTiO}_3$}
\newcommand{\PTO}{$\text{PbTiO}_3$}

\begin{document}

\title{Can we trust the temperature - misfit strain thin film phase diagrams?}

\author{Alexander Kvasov}
\author{Alexander K. Tagantsev}

\affiliation{%
Ceramics Laboratory\\
\'Ecole Polytechnique F\'ed\'erale de Lausanne \\
CH-1015 Lausanne, Switzerland
}%
\date{\today}

\begin{abstract}
Currently used methods for the description of thermodynamics of ferroelectric thin films (Landau theory or ab initio based Monte-Carlo simulations) are based on an energy expansion in terms of internal degrees of freedom.
It was shown that these methods can suffer from a substantial inaccuracy unless higher-order electromechanical interactions are not taken into account.
The high-order electromechanical couplings strongly renormalize the sixth-power polarization terms of the thermodynamic energy expansion.
In this paper, apart from the general statement, we illustrate it with an example of a temperature - misfit strain phase diagram of a \BTO{} thin film derived with high-order electromechanical interactions evaluated using first principles calculations.
\end{abstract}

\maketitle

\section{Introduction}\label{s_introduction}
Strain engineering is a modern strategy to control and enhance materials properties. It represents a technique which deals with semiconductor and ferroic thin films strained on a substrate.
Such strain can be tensile or compressive and occurs due to a difference between lattice parameters of the film and that of the underlying substrate.
In strain-engineered ferroics, the strain appearing in the film shifts the transition temperatures and can change the properties of the material such as the dielectric and piezoelectric constants, remanent polarization, or even can induce room temperature ferroelectricity in a non-ferroelectric material \cite{Haeni_Nature_2004}.
Currently used methods for the description of thermodynamics of ferroic thin films (classical Landau theory \cite{Pertsev_1998}, phase field modeling \cite{Li_2001}, and \textit{ab initio} based Monte-Carlo (MC) simulations \cite{Vanderbilt_BTO_2004}) are based on an energy expansion in terms of internal degrees of freedom.
Any treatment of this kind starts from such expansion for the bulk ferroic (e.g. the effective Hamiltonian for MC simulations \cite{Vanderbilt_BTO_2004} or a thermodynamic potential for Landau theory analysis \cite{Pertsev_1998}).
Further mixed mechanical conditions corresponding to the films are applied.
This way one establishes a kind of effective thermodynamic potential (or effective Hamiltonian) of the film.
Minimisation of such potential or further MC simulations yield the ferroic state of a film as a function of the temperature and misfit strain.
The standard way to present the results of such calculations is the so-called "temperature - misfit strain" phase diagrams \cite{Pertsev_1998}, which give  the ferroic state depending on temperature and strain appearing in the film.
Such diagrams have been developed for many classical ferroelectric materials like \BTO \cite{Pertsev_1998,Pertsev_Ferroelectrics_1999,Koukhar_PhysRevB_2001,Li_2001,Vanderbilt_BTO_2004}, \PTO \cite{Pertsev_Ferroelectrics_1999,Koukhar_PhysRevB_2001}, and \STO \cite{Pertsev_PhysRevB_2000}.

The goal of this work is to draw attention to the fact that the aforementioned methods of description of ferroelectric thin films in their current implementations may suffer from a serious \emph{principle}  drawback, which may lead to erroneous results.
It is commonly understood that a large (sometimes enormous) difference between the properties of a strained film and its bulk counterpart is due to the coupling between the order parameter and strain.
Customarily this coupling is modeled in the so-called electrostriction approximation corresponding to terms quadratic in the order parameter and linear in mechanical strain (stress) in the  thermodynamic potential \cite{Pertsev_1998} or effective Hamiltonian \cite{Vanderbilt_BTO_2004} of the bulk material.
Already 10 years ago, when using the electrostriction approximation for a special situation in \PTO{} films of (111)-orientation, a possible \emph{principle} deficiency of this approximation was pointed out.
This paper is devoted to a comprehensive analysis of this problem in terms of Landau theory however the conclusion should hold for MC simulations as well.
We show that this "electrostriction based" description provides an adequate thermodynamics approach of ferroelectric thin films only if it is controlled by the free energy expansion up to fourth power terms in polarization.
However, if the sixth-power polarization terms ($\gamma$-terms) are needed, which is a common situation for ferroelectric perovskites, higher-order electromechanical couplings should be taken into account.
We elucidate the matter in terms of a simplified phenomenological model.
To demonstrate the phenomenon for thin films, we estimate some higher-order electrostrictive coefficients using experimental data and we calculate missing coefficients (in order to have a complete set) using \textit{ab initio} methods. Then, we show that the higher-order electromechanical couplings readily lead to an order-of-magnitude renormalization of the $\gamma$-terms of free energy when passing from the bulk thermodynamic potential to the effective thermodynamic potential of the film.
Finally, we illustrate our message with the results of the calculations for a \BTO{} thin film.

\section{Scalar model}\label{s_scalar_model}
The problem with the "electrostriction based" description of strained ferroelectrics can be illustrated qualitatively by a simple scalar model.
Let us describe a ferroelectric with a Gibbs thermodynamic potential energy expansion keeping only one component of polarization $P$ and stress $\sigma$:
\begin{widetext}
\begin{equation}\label{G_bulk_scalar}
G=\underbrace{\frac{\alpha}{2} P^2 + \frac{\beta}{4} P^4 + \frac{\gamma}{6} P^6 - \frac{s}{2} \sigma^2 - Q P^2 \sigma}_\text{"ordinary"} - \underbrace{\frac{M}{2} P^2 \sigma^2 - R P^4 \sigma - \frac{N}{3} \sigma^3}_\text{"high-order"},
\end{equation}
\end{widetext}
where the "ordinary" part represents Gibbs energy commonly used to describe ferroelectric systems ($Q$ - "ordinary" electrostrictive coefficient, $s$ - linear compliance) and "high-order" terms ($M$, $R$ - high-order electrostrictive coefficients and $N$ - non-linear compliance) which are customarily neglected.
This neglect can be readily justified for a bulk material but the situation for a clamped system is different.
We can show this considering ferroelectricity in a clamped system, i.e. where the strain $\epsilon$ equals the misfit strain $\epsilon_0$.
To obtain an effective potential of the clamped system
\begin{equation}\label{eq_G_film}
\widetilde{G}(T,P,\epsilon_0)= G + \epsilon \sigma,
\end{equation}
where $T$ is temperature, we eliminate stress $\sigma$ using a mechanical equation of state
\begin{equation}\label{mech_eq}
\epsilon = - \frac{\partial{G}}{\partial{\sigma}} = s \sigma + N \sigma^2 + Q P^2 + R P^4 + M P^2 \sigma,
\end{equation}
and write the effective thermodynamic potential $\widetilde{G}(T,P,\epsilon_0) $:
\begin{eqnarray}\label{G_film_1D}
\widetilde{G} = \frac{\alpha^*}{2} P^2+\frac{\beta^*}{4} P^4+\frac{\gamma^*}{6} P^6+\frac{\epsilon^2_0}{2 s},
\end{eqnarray}
where minima of $\widetilde{G}$ with respect to polarization correspond to the ground state of the clamped system.
Electromechanical interactions in such systems lead to renormalizations of $\alpha$, $\beta$ and $\gamma$:
\begin{equation}\label{alpha_change}
\alpha^*=\alpha - \epsilon_0 \frac{Q}{s} - \underbrace{\epsilon_0^2 (\frac{M}{2 s^2} + \frac{Q N}{s^3})}_\text{A},
\end{equation}
\begin{equation}\label{beta_change}
\beta^*=\beta + 2\frac{Q^2}{s} + \underbrace{\epsilon_0 (\frac{M Q}{s^2} - \frac{R}{s} - \frac{N Q^2}{s^3})}_\text{B},
\end{equation}
\begin{equation}\label{gamma_change}
\gamma^*=\gamma - \underbrace{3 M \frac{Q^2}{s^2} + 6 R \frac{Q}{s} + 2 N \frac{Q^3}{s^3}}_\text{C}.
\end{equation}
Now, let us have a closer look at Eqs. \eqref{alpha_change}-\eqref{gamma_change}.
The $\epsilon_0 \frac{Q}{s}$ term in \eqref{alpha_change} leads to the shift of the phase transition temperature.
The $2\frac{Q^2}{s}$ term in \eqref{beta_change} renormalizes $\beta$, for example, for \BTO{} this "renormalization" switches the sign of $\beta$, i.e. it changes the order of the phase transition passing from a bulk material to a film \cite{Pertsev_Ferroelectrics_1999}.
These two "electrostrictive" corrections do not involve high-order electrostrictive couplings, they are well known and justified experimentally \cite{Choi_Science_2004,Haeni_Nature_2004}.
At the same time, atomic order-of-magnitude estimates show that, in \eqref{alpha_change}, $\frac{Q}{s}$ is about $(\frac{M}{2 s^2} + \frac{Q N}{s^3})$ whereas $2\frac{Q^2}{s}$ is about $(\frac{M Q}{s^2} - \frac{R}{s} - \frac{N Q^2}{s^3})$ in \eqref{beta_change}.
Thus, in view of smallness of $\epsilon_0$ for any practical situation the $A$ and $B$ corrections are expected to be negligible, except the cases where the low-order corrections are unusually small.
However, atomic order-of-magnitude estimates suggest that the $C$ correction in \eqref{gamma_change} is of the same order of magnitude as $\gamma$ similar to the strong renormalization of the $\beta$ term in \eqref{beta_change}.
These estimates imply that the high-order coefficients, which are customarily neglected in the majority of problems, should be included into thermodynamic energy expansions as far as the "$\gamma$-term" in \eqref{G_bulk_scalar} is important for the description of a problem.

\section{Landau theory of thin films with high-order \\ electrostrictive couplings}\label{s_thin_film_theory}
There are several ways to introduce high-order couplings to Landau theory. The first one is to use the Gibbs thermodynamic energy expansion for a centrosymmetric cubic crystal with respect to polarization $P_i$ and stress $\sigma_{ij}$:
\begin{eqnarray}\label{G_bulk}
G=a P^2 + a_{ij} P^2_i P^2_j+a_{ijk} P^2_i P^2_j P^2_k - \frac{s_{ijkl}}{2} \sigma_{ij} \sigma_{kl} -
\frac{N_{ijklmn}}{3} \sigma_{ij} \sigma_{kl} \sigma_{mn} - Q_{ijkl} P_i P_j \sigma_{kl} - \\ \nonumber
- \frac{M_{ijklmn}}{2} P_i P_j \sigma_{kl} \sigma_{mn} - R_{ijklmn} P_i P_j P_k P_l \sigma_{mn},
\end{eqnarray}
where $a$, $a_{ij}$, and $a_{ijk}$ are dielectric stiffness and higher-order stiffness coefficients at constant stress, $s_{ijkl}$ and $N_{ijklmn}$ are linear and nonlinear elastic compliances, $Q_{ijkl}$ is ordinary electrostriction, and $M_{ijklmn}$ and $R_{ijklmn}$ are high-order electrostriction tensors.
Hereafter we assume summation over repeated indices.
The minima of $G$ with respect to polarization correspond to the ground state of the mechanically free sample.
The $G$ expansion is often used dealing with experimental data.
Alternatively, when working with \emph{ab initio} calculations, instead of $G$ expansion one naturally can use the Helmholtz thermodynamic function $F$ written in terms of polarization $P_i$ and strain $\epsilon_{ij}$:
\begin{eqnarray}\label{F_bulk}
F=b P^2 + b_{ij} P^2_i P^2_j+b_{ijk} P^2_i P^2_j P^2_k+ \frac{c_{ijkl}}{2} \epsilon_{ij} \epsilon_{kl} +
\frac{n_{ijklmn}}{3} \epsilon_{ij} \epsilon_{kl} \epsilon_{mn} - q_{ijkl} P_i P_j \epsilon_{kl} - \\ \nonumber
- \frac{m_{ijklmn}}{2} P_i P_j \epsilon_{kl} \epsilon_{mn} - r_{ijklmn} P_i P_j P_k P_l \epsilon_{mn},
\end{eqnarray}
where $q_{ijkl}$, $m_{ijklmn}$ and $r_{ijklmn}$ are components of linear and high-order electrostrictive tensors, and $c_{ijkl}$ and $n_{ijklmn}$ are linear and non-linear stiffness tensors. The minima of $F$ correspond to the ground state of a fully mechanically clamped sample.

The relationships between the $N_{ijklmn}$ and $n_{ijklmn}$ coefficients can be found by resolving the mechanical state equations for stress:
\begin{equation}
\sigma_{ij} = c_{ijkl} \epsilon_{kl} + n_{ijklmn} \epsilon_{kl} \epsilon_{mn} \label{eq_n-N_sigma}
\end{equation}
and strain:
\begin{equation}
\epsilon_{ij} = s_{ijkl} \sigma_{kl} + N_{ijklmn} \sigma_{kl} \sigma_{mn}. \label{eq_n-N_u}
\end{equation}
Eliminating, for example, stress $\sigma_{ij}$ between \eqref{eq_n-N_sigma} and \eqref{eq_n-N_u} and keeping linear terms in \eqref{eq_n-N_sigma} in view of the smallness of $\epsilon_{ij}$ one can obtain:
\begin{equation}\label{eq_n-N}
n_{ijklmn} = - c_{ijuv} N_{uvwxyz} c_{klwx} c_{mnyz}.
\end{equation}

Hereafter, in addition to standard Voigt notations for stress $\sigma_i$, strain $\epsilon_i$, elastic ($s_{ij}$ and $c_{ij}$) \cite{Nye_book} and linear electrostriction ($q_{ij}$ and $Q_{ij}$) tensors, defining  $Q_{ij}$ according to the Landolt-Bornstein reference book \cite{Landolt-Bornstein}
\begin{equation} \label{eq_voigt_Q}
Q_{ijkl} =
\begin{cases}
Q_{mn} \text{ for } n=1,2,3 \\
\frac{Q_{mn}}{2} \text{ for } n=4,5,6
\end{cases}
,
\end{equation}
we use the Voigt matrix notation for nonlinear elasticity ($n_{ijk}$ and $N_{ijk}$) and high-order electrostriction ($m_{ijk}$, $r_{ijk}$, $M_{ijk}$, and $R_{ijk}$) tensors as follows:
\begin{equation} \label{eq_voigt_N}
N_{ijklmn} =
\begin{cases}
N_{abc}, ~ a,b,c=1,2,3 \\
\frac{N_{abc}}{2}, \text{ one suffix $4,5,6$ and other $1,2,3$} \\
\frac{N_{abc}}{4}, \text{ one suffix $1,2,3$ and other $4,5,6$} \\
\frac{N_{abc}}{8}, \text{ for } a,b,c=4,5,6
\end{cases}
,
\end{equation}
\begin{eqnarray} \label{eq_voigt_M}
\begin{cases}
M_{ijklmn} =M_{abc}, ~ a,b,c=1,2,3 \\
M_{ijklmn} =\frac{M_{abc}}{2}, ~ a=1,2,3; b,c=(\text{one suffix $4,5,6$ and another $1,2,3$}) \\
M_{ijklmn} =\frac{M_{abc}}{4}, ~ a=1,2,3; b,c=4,5,6 \\
M_{414}=8 M_{231123}, ~ M_{424}=8 M_{232223}, ~ M_{456} = 8 M_{231312}
\end{cases}
,
\end{eqnarray}
\begin{eqnarray} \label{eq_voigt_R}
\begin{cases}
R_{ijklmn} =R_{abc}, ~ a,b,c=1,2,3 \\
R_{ijklmn} =\frac{R_{abc}}{2}, ~ a,b=1,2,3;c=4,5,6 \\
R_{144} = 4 R_{112323}, ~ R_{155} = 4 R_{111313}
\end{cases}
.
\end{eqnarray}
The Voigt matrix notation for the coefficients of the $F$ expansion are $n_{ijklmn} = n_{abc}, ~ a,b,c=1..6$; $m_{ijklmn} = m_{abc}, ~ a=1,2,3;b,c=1..6$ and $m_{414}=2 m_{231123}, m_{424}=2 m_{232223}, m_{456} = 2 m_{231312}$; and $r_{ijklmn} = r_{abc}, ~ a,b=1,2,3;c=1..6$ and $r_{144} = 2 r_{112323}, r_{155} = 2 r_{111313}$.

Knowing the $q_{ij}$, $m_{ijk}$, and $r_{ijk}$ coefficients of the $F$ expansion one can find the corresponding $Q_{ij}$, $M_{ijk}$, and $R_{ijk}$ coefficients for expansion $G$ since both thermodynamic functions correspond to the same state equation.
For example, one can find for $M_{111}$:
\begin{eqnarray}\label{eq_m111_transform}
M_{111}=\frac{\left(c_{11}+c_{12}\right){}^2 m_{111}}{\left(c_{11}-c_{12}\right){}^2 \left(c_{11}+2 c_{12}\right){}^2}-
\frac{4c_{12} \left(c_{11}+c_{12}\right) m_{112}}{\left(c_{11}-c_{12}\right){}^2 \left(c_{11}+2 c_{12}\right){}^2}+ \\ \nonumber
\frac{2 c_{12}^2 m_{122}}{\left(c_{11}-c_{12}\right){}^2
\left(c_{11}+2 c_{12}\right){}^2}+
\frac{2 c_{12}^2 m_{123}}{\left(c_{11}-c_{12}\right){}^2 \left(c_{11}+2 c_{12}\right){}^2}.
\end{eqnarray}
Other relationships between coefficients of $G$ and $F$, in view of their complexity, are presented in Appendix \ref{app_legendre_transform}.

To illustrate quantitatively the phenomenon described above in Sec. \ref{s_scalar_model}, we exploit an effective energy potential of a strained thin film of a single-domain (001)-oriented \BTO{}.
To obtain the effective potential from the $G$ expansion:
\begin{equation}\label{G_eff}
\widetilde{G}(P_i,T,\epsilon_0) = G+\epsilon_1 \sigma_1+\epsilon_2 \sigma_2+\epsilon_6 \sigma_6,
\end{equation}
the minima of which correspond to the ground state of the film partially clamped on the substrate \cite{Pertsev_Ferroelectrics_1999}, we apply mixed mechanical conditions
\begin{eqnarray}\label{eq_boundary_conditions}
\frac{\partial G}{\partial \sigma_1}=\epsilon_0,\frac{\partial G}{\partial \sigma_2}=\epsilon_0,\frac{\partial G}{\partial \sigma_6}=0 \\
\sigma_3=0,\sigma_4=0,\sigma_5=0 \nonumber
\end{eqnarray}
and eliminate stresses $\sigma_1$, $\sigma_2$ and $\sigma_6$.
Hereafter a Cartesian coordinate system with the $x_3$ axis perpendicular to the film-substrate interface is considered, $\epsilon_0=\frac{a_\parallel-a_0}{a_0}$ - biaxial parent misfit strain, where $a_0$ is the lattice parameter of the ferroelectric material in the cubic phase extrapolated to temperature $T$ and $a_\parallel$ is the in-plane lattice parameter of the film. Minimizing $\widetilde{G}(P_i,T,\epsilon_0)$ with respect to polarization $P_i$ for each $(T,\epsilon_0)$ point one can find the polarization states of the thin film, i.e. one can build the phase diagram.

As it was mentioned in discussing the scalar model, the high-order electrostrictive as well as nonlinear compliance coefficients lead to the changes in coefficients $a_{ijk}^\text{bulk} \rightarrow a_{ijk}^\text{film}$ of $P^6$-terms while passing from a bulk crystal to the thin film ($G \rightarrow \widetilde{G} $).
Here, we give an example of such a change for $a^\text{film}_{333}$:
\begin{eqnarray}\label{eq_a333_transform}
a^\text{film}_{333} = a_{111}^\text{bulk} - \frac{M_{122} Q_{12}^2}{\left(s_{11}+s_{12}\right){}^2}-\frac{M_{123} Q_{12}^2}{\left(s_{11}+s_{12}\right){}^2}+\frac{2 Q_{12}
R_{112}}{s_{11}+s_{12}}+ \frac{2 Q_{12}{}^3 \left(N_{111}+3 N_{112}\right)}{3 \left(s_{11}+s_{12}\right){}^3}.
\end{eqnarray}
The other exact expressions of the $a_{ijk}^\text{bulk} \rightarrow a_{ijk}^\text{film}$ change can be found in Appendix \ref{app_bulk_film}.

\section{Ab initio calculations}\label{s_ab_initio}
To see what impact high-order interactions exert on the thin film effective potential $\widetilde{G}$ \eqref{G_eff}, one has to know the values of the high-order electrostrictive and non-linear compliance coefficients.
Since the experimental information on the high-order coefficients is scarce we turned towards \emph{ab initio} methods, namely, we used the Vienna Ab-initio Simulation Package (VASP) \cite{VASP_Kresse_1996} performing zero Kelvin Density Functional Theory (DFT) full relaxation calculations. All calculations were performed within the generalized-gradient approximation as implemented in VASP using the projector augmented-wave method for the electron-ion interactions \cite{VASP_Kresse_1999}. We have used a 8x8x8 Monkhorst-Pack grid for k-point sampling \cite{Monkhorst_1976}, and a plane-wave energy cut-off of 600eV. For full relaxation calculations, the threshold of the Hellman-Feynman force was less than 1meV/A.
We would like to underline that mechanical compliance and electrostriction in \eqref{G_bulk} are expected to be weakly temperature dependent. This justifies the use of zero Kelvin DFT results in finite temperature calculations.

Technically, because of working with DFT, it is more convenient to use the $F$ expansion \eqref{F_bulk} to calculate the high-order coefficients.
Stiffness $c_{ij}$ and $n_{ijk}$ and electrostrictive $q_{ij}$, $m_{ijk}$, and $r_{ijk}$ coefficients can be found in the following way.
Using VASP we find stress $\sigma_i$ on strain $\epsilon_j$ and polarization $P_i$ on strain $\epsilon_j$ dependences for different $i$ and $j$, then, we use the mechanical state equation
\begin{equation}\label{mech_eq_high}
\sigma_{ij}= \frac{\partial G}{\partial \epsilon_{ij}} = c_{ijkl} \epsilon_{kl} + n_{ijklmn} \epsilon_{kl} \epsilon_{mn} - q_{klij} P_k P_l - m_{klmnij} P_k P_l \epsilon_{mn} - r_{klmnij} P_k P_l P_m P_n,
\end{equation}
and determine the coefficients by fitting.
Polarization was calculated by the atomic displacements $\xi_i$ and Born charges for the cubic phase $Z_i$:
\begin{equation}\label{eq_p}
P = \frac{e}{V_0} \sum_i Z_i \xi_i,
\end{equation}
where $i$ enumerates the atoms in the unit cell, $e$ - the charge of electron, $V_0$ - volume of the cubic unit cell.

\begin{figure}[!ht]
\includegraphics[width=0.48\textwidth]{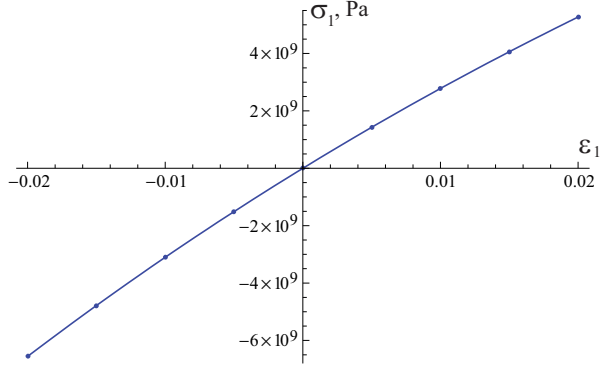}
\caption{Modeled stress $\sigma_1$ on strain $\epsilon_1$ dependence for \BTO{}. Obtained with VASP keeping the $mmm$ symmetry of the structure.}
\label{fig_u}
\end{figure}

It is possible to find $c_{ij}$ and $n_{ijk}$ tensors separately from others ($q_{ij}$, $m_{ijk}$, and $r_{ijk}$) if one uses the mechanical state equation \eqref{mech_eq_high} at zero polarization and models the deformation applied to the paraelectric cubic phase keeping the $mmm$ symmetry of the structure.
For example, the mechanical state equation \eqref{mech_eq_high} at zero polarization and only $\epsilon_1$ nonzero component gives:
\begin{equation}\label{mech_eq_u1}
\sigma_1=c_{11} \epsilon_1 + n_{111} \epsilon^2_1.
\end{equation}
The $\sigma_1(\epsilon_1)$ dependence obtained with VASP shown in Fig. \ref{fig_u} are inserted into Eq. \eqref{mech_eq_u1} and the corresponding $c_{11}$ and $n_{111}$ coefficients are obtained by fitting.

Then, the other $q_{ij}$, $m_{ijk}$, and $r_{ijk}$ coefficients can be found from analysis of all of Eq. \eqref{mech_eq_high}.
Let us demonstrate how to find the $q_{11}$, $m_{111}$ and $r_{111}$ tensors components.
One uses Eq. \eqref{mech_eq_high} where only polarization $P_1$ and strain $\epsilon_1$ are nonzero:
\begin{equation}\label{mech_eq_111}
\sigma_1=c_{11} \epsilon_1 + n_{111} \epsilon^2_1 - q_{11} P^2_1 - m_{111} P^2_1 \epsilon_1 - r_{111} P^4_1,
\end{equation}
Using VASP full relaxation calculations and keeping the $4mm$ tetragonal symmetry of the structure, we find the $P_1(\epsilon_1)$ and $\sigma_1(\epsilon_1)$ dependences and substitute them into Eq. \eqref{mech_eq_111}.
\begin{figure}[!ht]
\includegraphics[width=0.48\textwidth]{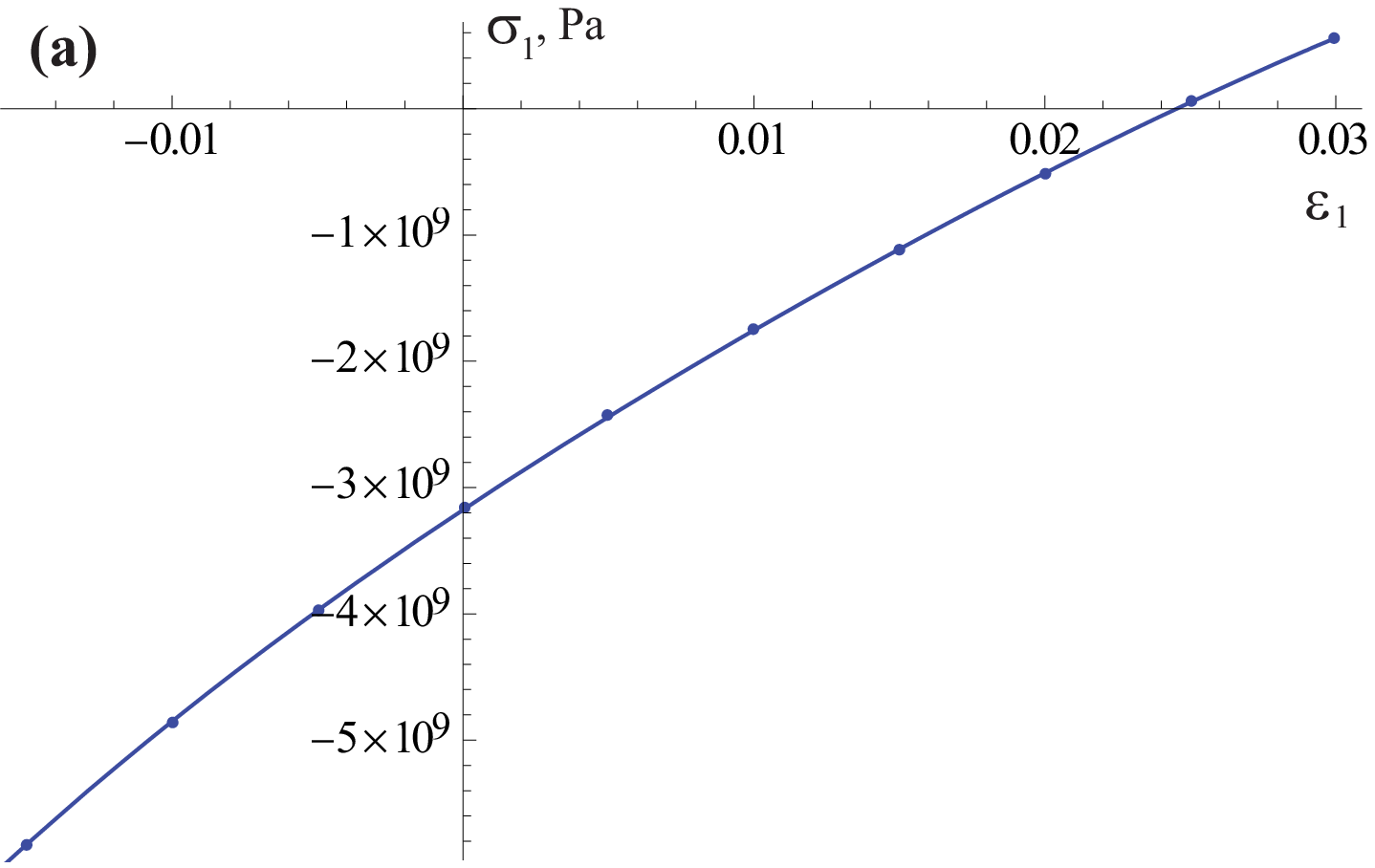}
\includegraphics[width=0.48\textwidth]{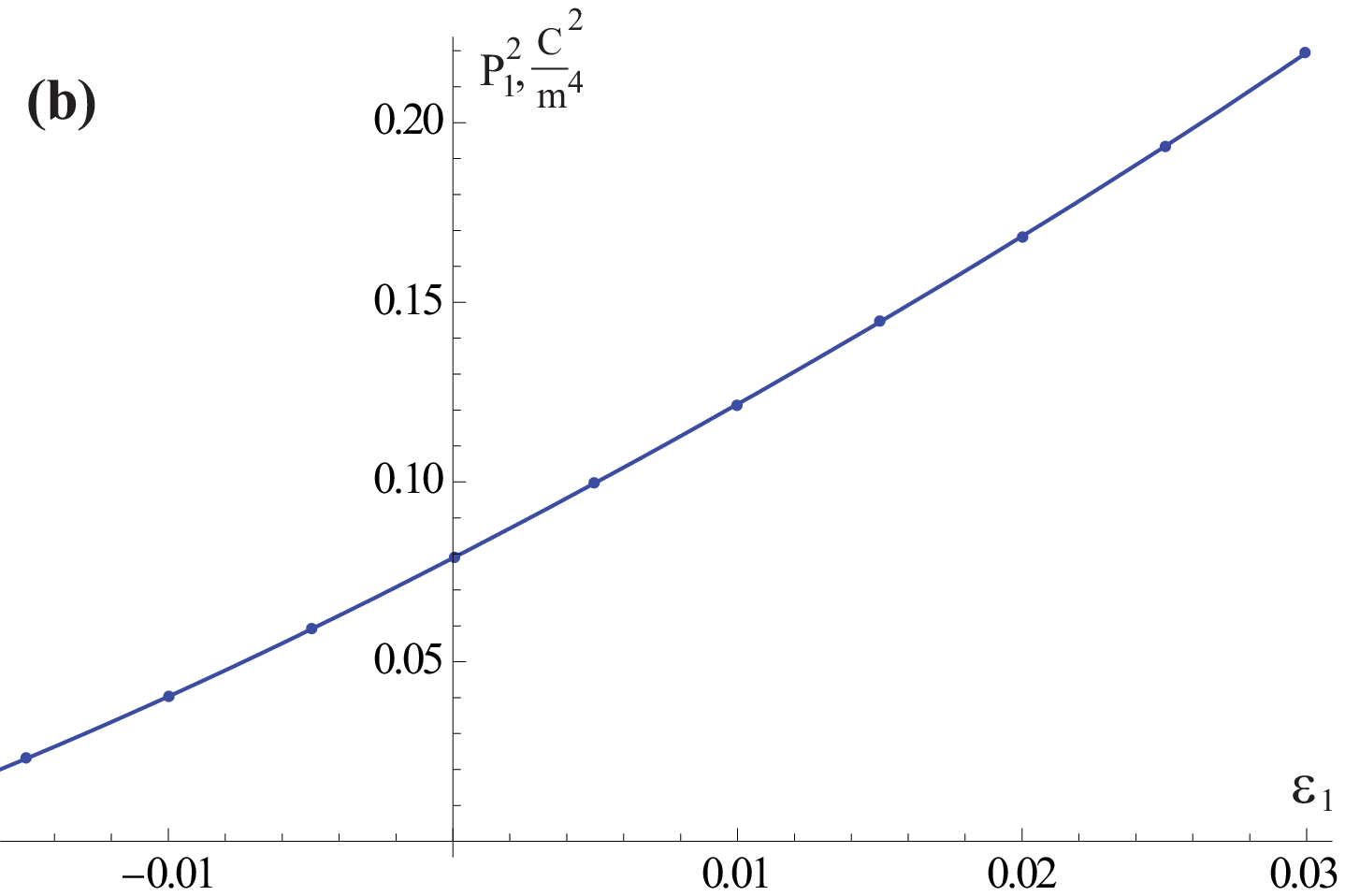}
\caption{Modeled stress $\sigma_1$ on strain $\epsilon_1$ (a) and square of polarization $P_1^2$ on strain $\epsilon_1$ (b) dependences for \BTO{}. Obtained with VASP full relaxation calculations keeping the $4mm$ tetragonal symmetry of the structure.}
\label{fig_P_u}
\end{figure}
Once we have polynomial functions of $\epsilon_1$ on the both sides of Eq. \eqref{mech_eq_111} we can in principle obtain $q_{11}$, $m_{111}$ and $r_{111}$ by fitting.
By repetition of such a procedure for different symmetries of the structure and for different components of $\sigma_i$ and $\epsilon_j$ it is possible to calculate all the components of the high-order electrostrictive tensors.

The above described scheme requires a good accuracy of $\sigma_i(\epsilon_j)$ and $P_i(\epsilon_j)$ dependences. Unfortunately, because of the technical limitations of software we can not have reliable precision of the curvature of $\sigma_i(\epsilon_j)$ and $P_i(\epsilon_j)$ and therefore reliable precision of $r_{ijk}$ coefficients.
We clarify the matter with the example where only $\sigma_1(\epsilon_1)$ and $P_1(\epsilon_1)$ are nonzero.
One represents obtained with VASP $\sigma_1(\epsilon_1)$ and $P_1^2(\epsilon_1)$ dependences as
\begin{eqnarray}\label{mech_eq_P_u_expan}
\sigma_1(0) = \sigma_1(0) + \sigma_1'(0) \epsilon_1 + \sigma_1''(0) \epsilon_1^2, \\
P_1^2(0) = \rho_0 + \rho_1 \epsilon_1 + \rho_2 \epsilon_1^2
\end{eqnarray}
and inserts them into Eq. \eqref{mech_eq_111}.
Equating coefficients of powers of $\epsilon_1$ one has 3 independent equations:
\begin{eqnarray}
\sigma_1(0)= - q_{11} \rho_0 - r_{111} \rho_0^2, \label{mech_eq_expan_1} \\
\sigma_1'(0) = c_{11} - q_{11} \rho_1 - 2 m_{111} \rho_0 - 2 r_{111} \rho_0 \rho_1, \label{mech_eq_expan_2} \\
\sigma_1''(0) = n_{111} - q_{11} \rho_2 - 2 m_{111} \rho_1 - r_{111} (\rho_1^2 + 2 \rho_0 \rho_2) \label{mech_eq_expan_3}
\end{eqnarray}
to find simultaneously 3 unknown values of $q_{11}$, $m_{111}$, and $r_{111}$.
This way one has to consider the second derivatives of $\sigma_1(\epsilon_1)$ and $P_1(\epsilon_1)$ dependences (Eq. \eqref{mech_eq_expan_3}) which cannot be reliably found with VASP.
This is supported by the fact that the $r_{ijk}$ values found with VASP are in conflict with those obtained with experimental data.
Thus, we do not attempt to calculate the $r_{ijk}$ coefficients, and this way we use only the two lower-order independent equations.
We realize that the exclusion of $r_{ijk}$ can cause the change of $m_{ijk}$ in \eqref{mech_eq_expan_2}, but the numerical calculations show that this leads to a small correction of $m_{ijk}$.

The coefficients are calculated with errors as it can be seen from Table \ref{table_VASP_coeff}. The origin of the error of the coefficients is due to the computational limit of VASP causing the calculation error of atomic displacements and therefore polarization (which was calculated by atomic displacements) and stress $\sigma$. Additionally the above described calculation scheme is iterative, one needs to use previously found coefficients to calculate new ones. For example, to find a coefficient $m_{112}$ which corresponds to the term
\begin{equation}\label{eq_F_for_m}
F=...- \frac{m_{112}}{2} P_1^2 \epsilon_1 \epsilon_2 + ...
\end{equation}
one uses the mechanical equation of state keeping only $P_1$, $\epsilon_1$ and $\epsilon_2$ components:
\begin{equation}\label{eq_state_eq_for_m}
\frac{\partial F}{\partial \epsilon_1} = \sigma_1 = c_{11} \epsilon_1 + c_{12} \epsilon_2 + c_{111} \epsilon_1^2 +
2 c_{112} \epsilon_1 \epsilon_2 - q_{11} P_1^2 - m_{111} P_1^2 \epsilon_1 - m_{112} P_1^2 \epsilon_2.
\end{equation}
Then, using VASP we model stress and polarization, keeping $\epsilon_1 = \epsilon_2 = \epsilon$:
\begin{equation}\label{eq_state_eq_for_m2}
\sigma_1(\epsilon) = (c_{11} + c_{12}) \epsilon + (c_{111} + 2 c_{112}) \epsilon - q_{11} P_1^2 - (m_{111} + m_{112}) P_1^2 \epsilon.
\end{equation}
The $\sigma_1(\epsilon)$ and $P_1(\epsilon)$ dependences obtained with VASP have some errors as well as the previously found $c_{11}$, $c_{12}$, $c_{111}$, $c_{112}$, $q_{11}$, and $m_{111}$ coefficients. So, when the $m_{112}$ coefficient is calculated it contains an accumulating error of all previously found coefficients and clearly has lower precision then for example $m_{111}$.

\section{Experimental estimates of $R_{ijk}$ coefficients}\label{s_R}
Due to the technical limitations of VASP we are unable to reliably determine the $r_{ijk}$ coefficients.
Luckily we can estimate some of the $R_{ijk}$ coefficients using experimental information on piezoelectric coefficients and spontaneous polarization and stain.
From \eqref{G_bulk} one can proceed to the linearized constitutive equations for the piezoelectric coefficient $d_{ij}$ and spontaneous strain $\epsilon_S$. For \BTO{} in the tetragonal phase we have:
\begin{eqnarray}
d_{33} = \chi_{33} \frac{\partial}{\partial P_3} \left( \frac{\partial G}{\partial \sigma_3} \right) = 2 Q_{11} P_S \chi_{33} + 4 R_{111} P^3_S \chi_{33} \label{eq_d33} \\
\epsilon_{S3} = \frac{\partial G}{\partial \sigma_3} = Q_{11} P_S^2 + R_{111} P_S^4 \label{eq_u33} \\
d_{31} = \chi_{33} \frac{\partial}{\partial P_3} \left( \frac{\partial G}{\partial \sigma_1} \right) = 2 Q_{12} P_S \chi_{33} + 4 R_{112} P^3_S \chi_{33} \label{eq_d31} \\
\epsilon_{S1} = \frac{\partial G}{\partial \sigma_3} = Q_{12} P_S^2 + R_{112} P_S^4 \label{eq_u11} \\
d_{15} = \chi_{11} \frac{\partial}{\partial P_1} \left( \frac{\partial G}{\partial \sigma_5} \right) = 2 P_S Q_{44} \chi_{11} + 4 R_{155} P^3_S \chi_{11}, \label{eq_d15}
\end{eqnarray}
where $P_S = 0.26 \frac{\text{C}}{\text{m}^2}$ \cite{Merz_1949} is the spontaneous polarization, $\chi_{33} = 168 \epsilon_0$ and $\chi_{11}=2920 \epsilon_0$ \cite{Berlincourt_1958} are the dielectric susceptibilities.
Eqs. \eqref{eq_d33}-\eqref{eq_d15} can be appended with the equation for the spontaneous stain in the orthorhombic phase:
\begin{equation}\label{eq_u5}
\epsilon_{S5} = \frac{\partial G}{\partial \sigma_5} = Q_{44} \frac{P_S^2}{2} + R_{155} \frac{P_S^4}{4},
\end{equation}
where for $P_S$ we take the same value of $0.26 \frac{\text{C}}{\text{m}^2}$ since the absolute value of $P_S$ does not change during the phase transition from the tetragonal to the orthorhombic phase \cite{Merz_1949}, there is only a rotation of polarization vector.
Resolving Eqs. \eqref{eq_d33}-\eqref{eq_u5} and using experimental data on $d_{ij}$ and $\epsilon_{Si}$ from Table \ref{table_Q_and_R} one can estimate the $Q_{ij}$ and $R_{ijk}$ values which are also shown in Table \ref{table_Q_and_R}.
\begin{table}[!ht]
\begin{tabular}{|c|c|c|c|}
\hline
$d_{ij}, \left[ 10^{-12}\frac{\text{C}}{\text{N}} \right]$ & $\epsilon_{Si}$ & $Q_{ij}, \left[\frac{\textrm{m}^4}{\textrm{C}^2}\right]$ & $R_{ijk},\left[\frac{\textrm{m}^8}{\textrm{C}^4}\right]$ \\
\hline
$d_{33} = 85.6$ & $\epsilon_{S3} = 0.0077$ & $Q_{11}=0.118$ & $R_{111}=-0.08$ \\
\hline
$d_{31}=-34.5$ & $\epsilon_{S1}=-0.0027$ & $Q_{12}=-0.036$ & $R_{112}=-0.07$ \\
\hline
$d_{15}=240$ & $\epsilon_{S5}=0.00105$ & $Q_{44}=0.032$ & $R_{155}=-0.02$ \\
\hline
\end{tabular}
\caption{Some experimental material parameters of \BTO{}. $d_{ij}$ - piezoelectric coefficients, the values are taken from Ref. \citenum{Berlincourt_1958}. $\epsilon_{Si}$ is the spontaneous strain, $\epsilon_{S3}=\frac{c-a_0}{a_0}$ and $\epsilon_{S1}=\frac{a-a_0}{a_0}$, where $c=4.034$ \AA{} and $a=3.992$ \AA{} are the lattice parameters of the tetragonal cell, $a_0=4.003$ \AA{} is the lattice constant of cubic \BTO{} extrapolated to room temperature \cite{Kay_1949}, $\epsilon_{S5}$ is recalculated from the distortion angle $\beta=89^{\circ} 51.6'$\cite{Kay_1949} of orthorhombic cell. $Q_{ij}$ and $R_{ijk}$ are linear and non-linear electrostriction calculated from experimental data.}
\label{table_Q_and_R}
\end{table}
Thus, 3 out of 6 components of the $R_{ijk}$ tensor can be found.

\section{Results and discussion}\label{s_results}
The coefficients of ordinary and high-order electrostriction as well as linear and non-linear elastic compliances obtained with the first principles calculations for the $F$ expansion were recalculated for $G$.
The analytic expressions of the recalculations can be found in Appendix \ref{app_legendre_transform} and \eqref{eq_n-N}.
The values of recalculated coefficients are given in Table \ref{table_VASP_coeff}.
\begin{table*}[!ht]
\centering
\begin{tabular}{|c|c|c|c|c|c|c|c|}
\hline
\multicolumn{2}{|c|}{$s_{ij},\left[10^{-12}\frac{1}{\textrm{Pa}}\right]$} & \multicolumn{2}{c|}{$s_{ij}^\text{exp},\left[10^{-12}\frac{1}{\textrm{Pa}}\right]$} &
\multicolumn{2}{c|}{$M_{ijk},\left[10^{-12}\frac{\textrm{m}^4}{\textrm{C}^2 \textrm{Pa}}\right]$} & \multicolumn{2}{c|}{$N_{ijk},\left[10^{-23}\frac{1}{\textrm{Pa}^2}\right]$}\\
\hline
$s_{11}$ & $4.25 \pm 0.01$ & $s_{11}$ & $8.3$ & $M_{111}$ & $-2.6 \pm 0.2$ & $N_{111}$ & $8 \pm 1$ \\
\hline
$s_{12}$ & $-1.14 \pm 0.01$ & $s_{12}$ & $-2.7$ & $M_{112}$ & $1.3 \pm 0.5$ & $N_{112}$ & $<2$ \\
\hline
$s_{44}$ & $8.45 \pm 0.01$ & $s_{44}$ & $9.3$ & $M_{122}$ & $3.5 \pm 0.5$ & $N_{123}$ & $<2$ \\
\hline
\multicolumn{4}{|c|}{} & $M_{123}$ & $-1.0 \pm 0.5$ & $N_{144}$ & $<2$ \\
\hline
\multicolumn{2}{|c|}{$Q_{ij},\left[\frac{\textrm{m}^4}{\textrm{C}^2}\right]$} &
\multicolumn{2}{c|}{$Q_{ij}^\text{exp},\left[\frac{\textrm{m}^4}{\textrm{C}^2}\right]$} & $M_{144}$ & $<1$ & $N_{155}$ & $<2$ \\
\hline
$Q_{11}$ & $0.162 \pm 0.005$ & $Q_{11}$ & $0.11$ & $M_{155}$ & $1.5 \pm 1$ & $N_{456}$ & $<4$ \\
\hline
$Q_{12}$ & $-0.034 \pm 0.005$ & $Q_{12}$ & $-0.043$ & $M_{414}$ & $<1$ & & \\
\hline
$Q_{44}$ & $0.021 \pm 0.005$ & $Q_{44}$ & $0.029$ & $M_{424}$ & $<1$ & & \\
\hline
 & & & & $M_{456}$ & $<1$ & & \\
\hline
\end{tabular}
\caption{Some material parameters of \BTO{} obtained from \emph{ab initio} calculations. $s_{ij}$ and $N_{ijk}$ are linear and nonlinear elastic compliance, $Q_{ij}$ and $M_{ijk}$ are linear and high-order electrostrictive tensors respectively. Experimental values of $s_{ij}$ and $Q_{ij}$ taken from Ref. \citenum{Pertsev_Ferroelectrics_1999} are also given for comparison.}
\label{table_VASP_coeff}
\end{table*}
We should note that current DFT methods give relatively moderate precision for high-order coefficients as it is clear from the error bars in Table \ref{table_VASP_coeff} due to technical limitations of the software and due to the neglect of the $r_{ijk}$ coefficients in the used scheme as it was described above.

Table \ref{table_gamma} demonstrates the size of the $a_{ijk}^\text{bulk} \rightarrow a_{ijk}^\text{film}$ renormalization effect for \BTO{} where the components of the original ($a_{ijk}^\text{bulk}$) and renormalized ($a_{ijk}^\text{film}$) tensors as well as the renormalizing corrections ($\Delta a_{ijk}$) are given.
The renormalized $a_{ijk}^\text{film}$ coefficients are obtained using relationships presented in Appendix \ref{app_bulk_film}.
\begin{table}[!ht]
\centering
\begin{tabular}{|c|c|c|c|c|}
\hline
 & $a_{ijk}^\text{bulk}$ & $\Delta a_{ijk}^\text{by $M_{ijk}$\&$R_{ijk}$}$ & $\Delta a_{ijk}^\text{by $N_{ijk}$}$ & $a_{ijk}^\text{film}$ \\
\hline
$a_{111}$ & 7.9 & $0.5 \pm 0.5$ & $1.4 \pm 0.2$ & $9.8 \pm 0.5$ \\
$a_{333}$ &     & $0.8 \pm 0.5$ & $< 0.2$ & $8.7 \pm 0.5$ \\
\hline
$a_{112}$ & 4.5 & $-10.6 \pm 0.5$ & $1.5 \pm 0.2$  & $-4.6 \pm 0.5$\\
$a_{113}$ &     & $-5.0 \pm 0.5$  & $-1.3 \pm 0.2$ & $-1.8 \pm 0.5$ \\
$a_{133}$ &     & $0.5 \pm 0.5$   & $0.5 \pm 0.2$  & $5.5 \pm 0.5$ \\
\hline
$a_{123}$ & 4.9 & $7.1 \pm 0.5$ & $1.4 \pm 0.2$ & $13.4 \pm 0.5$ \\
\hline
\end{tabular}
\caption{Renormalization of the coefficients of the $P^6$-terms when passing from the thermodynamic potential of a bulk material $G$ to the effective potential of a film $\widetilde{G}$.
All values are given in $10^{9}\frac{\textrm{m}^9}{\textrm{C}^4\textrm{F}}$ at 300K.
$a_{ijk}^\text{bulk}$ are coefficients of the expansion $G$ for bulk and mechanically free \BTO{}. $\Delta a_{ijk}$ are corrections to corresponding $a^\text{bulk}_{ijk}$ coefficients representing an addition to $a_{ijk}^\text{bulk}$ (for example, $a_{333}^\text{film}=a_{111}^\text{bulk} + \Delta a_{333}$). $a_{ijk}^\text{film}$ are coefficients of the $\widetilde{G}$ energy of the $(001)$-oriented clamped film.}
\label{table_gamma}
\end{table}
An inspection of this table shows that the renormalization is strong, e.g. $a_{123}^\text{bulk}=4.9 \times 10^{9}\frac{\textrm{m}^9}{\textrm{C}^4\textrm{F}}$ while $a_{123}^\text{film}=13.9 \times 10^{9}\frac{\textrm{m}^9}{\textrm{C}^4\textrm{F}}$ which means more than a 100\% change, confirming the conclusion drawn above from the order-of-magnitude estimates.

\begin{figure}[!ht]
\includegraphics[width=0.48\textwidth]{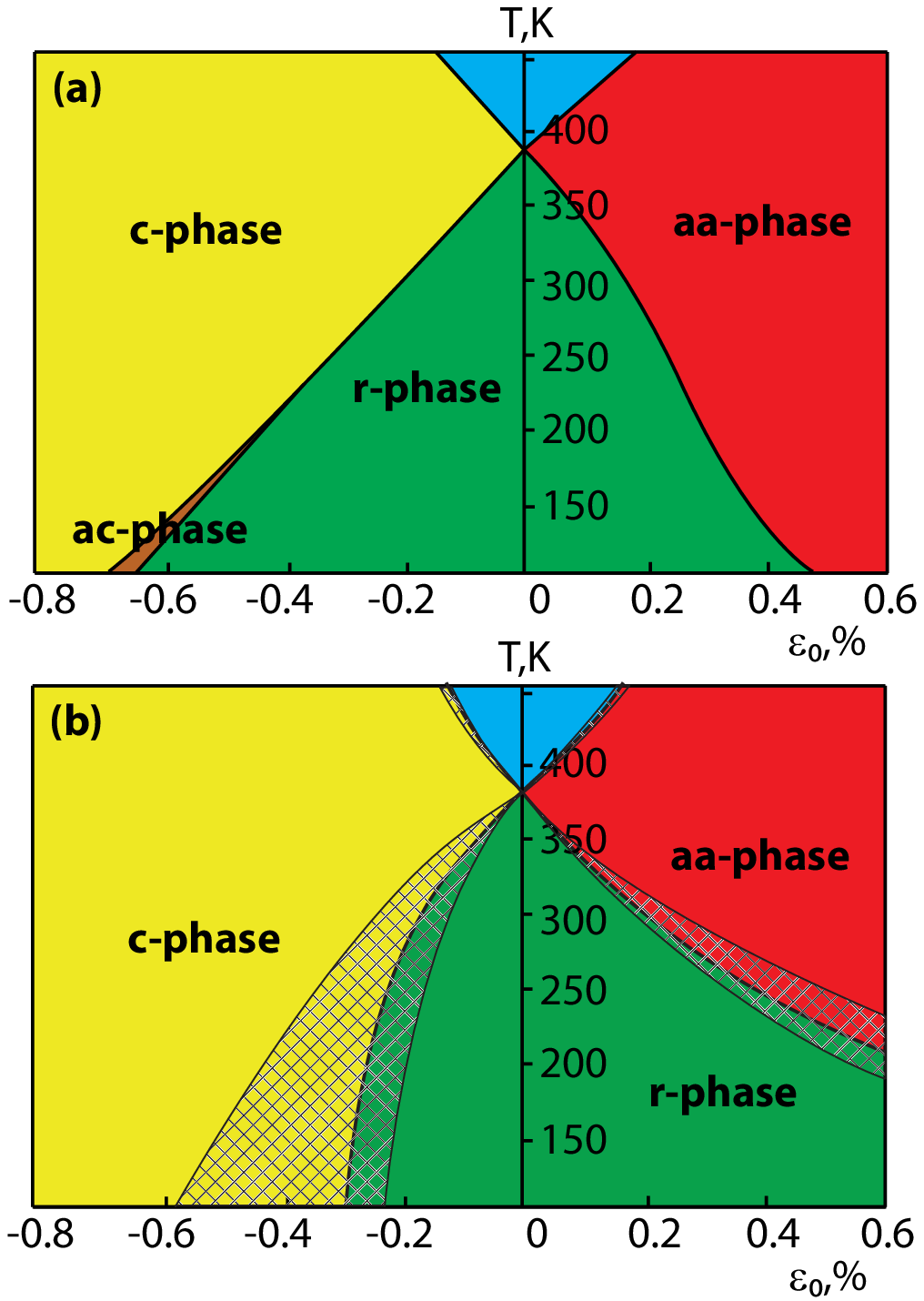}
\caption{Temperature - misfit strain phase diagrams of a single-domain (001)-oriented \BTO{}.
$\epsilon_0$ - biaxial parent misfit strain, T - temperature.
The phases are denoted as (i) the c-phase (yellow), polarization is out-of-plane ($P_1=P_2=0,P_3\neq0$); (ii) the aa-phase (red), where polarization is in-plane ($P_1\neq0,P_2\neq0,P_3=0$); (iii) the ac-phase (brown) where $P_1\neq0,P_2=0,P_3\neq0$; (iv) r-phase (red), where all components of polarization are non-zero; and (v) paraelectric phase (blue), polarization is 0.
(a) Original Pertsev's diagrams built with coefficients from  Ref. \citenum{Pertsev_Ferroelectrics_1999}.
(b) Developed with coefficients set from Ref. \citenum{Pertsev_Ferroelectrics_1999} appended with high-order $M_{ijklmn}$, $R_{ijklmn}$ and $N_{ijklmn}$ coefficients. The hatched regions demonstrate the shift of the transition lines within the error bars of the coefficients.
}
\label{fig_diagrams}
\end{figure}

Further we demonstrate the influence of high-order terms on the temperature - misfit strain phase diagram for a single-domain (001)-oriented \BTO{} thin film.
To plot the diagram, one minimizes the above obtained thin film effective potential $\widetilde{G}(P_i,T,\epsilon_0)$ \eqref{G_eff} with respect to polarization $P_i$ for each $(T,\epsilon_0)$ point.
A comparison between Figs. \ref{fig_diagrams}(a) and \ref{fig_diagrams}(b) shows how the original phase diagram built with the coefficients from Ref. \citenum{Pertsev_Ferroelectrics_1999} changes when supplemented with the high-order electrostrictive and nonlinear compliance coefficients taken from our first principles calculations (Table \ref{table_VASP_coeff}). From Figs. \ref{fig_diagrams}(a) and \ref{fig_diagrams}(b) one can see that taking into account the high-order coefficients strongly changes the diagram.
The effect might have been much stronger, but positive $P^4$-terms of $\widetilde{G}$ leading to the second-order phase transition in the film \cite{Pertsev_Ferroelectrics_1999} diminish the role of the strongly renormalized $a_{ijk}$ coefficients.
The hatched regions in Fig. \ref{fig_diagrams}(b) demonstrate the shift of the transition lines within the error bars of the coefficients. It is seen that the diagram is extremely sensitive to the variation of the high-order coefficients.

Another striking feature of the data from Table \ref{table_gamma} is the negative sign of some of $a_{ijk}^\text{film}$, making the renormalized theory formally unstable.
However, we found that physically the situation can be treated as stable.
The point is that the potential $\widetilde{G}$ of the thin film still has local minima due to the positive sign of renormalized coefficients for $P^4$-terms of the thermodynamic potential of the film \cite{Pertsev_Ferroelectrics_1999}, therefore the system is locally stable.
As for the global stability, it can be restored by adding $P^8$-terms in the $G$ expansion \cite{Wang_Tagantsev_2007}.
We do not incorporate these terms in our consideration, however we believe that this will not essentially affect the positions of the local minima in view of very high power of the terms.

We would like to note that the above results have implications on finite temperature MC simulations. To perform MC simulations for perovskites one customarily uses an effective Hamiltonian incorporating ordinary electrostriction \cite{Vanderbilt_BTO_2004,Zhong_1994}. In view of our findings, we suggest that for the treatment of ferroelectric thin films the effective Hamiltonian has to also include high-order electromechanical interactions and nonlinear elasticity.

\section{Conclusions}\label{s_conclusions}
To summarize, it was shown that an adequate Landau theory treatment of thermodynamics of typical ferroelectric thin films requires taking into account high-order electromechanical couplings and non-linear elasticity.
Our analysis also suggests that the \emph{ab initio} based Monte Carlo simulation of ferroelectric thin films involving the effective Hamiltonian should take into account not only customarily incorporated "ordinary electrostriction type" coupling, but also the high-order electromechanical interactions.
In view of this finding, we believe that an experimental evaluation of high-order electromechanical couplings in ferroelectrics seems to be a task of primary importance.

\begin{acknowledgments}
This work was supported by the Swiss National Science Foundation.
\end{acknowledgments}

\appendix
\section{Legendre transformation of high-order electrostrictive coefficients}\label{app_legendre_transform}
Knowing the coefficients of Helmholtz thermodynamic function $F$ written in terms of polarization $P_i$ and strain $\epsilon_i$ \eqref{F_bulk} it is possible to find the corresponding coefficients of Gibbs energy expansion $G$ with respect to polarization $P_i$ and stress $\sigma_i$ \eqref{G_bulk}.
Like in the main text we use the Voigt matrix notation for stress $\sigma_i$, strain $\epsilon_i$, and all elastic ($c_{ij}$, $n_{ijk}$, $s_{ij}$, and $N_{ijk}$) and electrostrictive ($q_{ij}$, $m_{ijk}$, $r_{ijk}$, $Q_{ij}$, $M_{ijk}$, and $R_{ijk}$) tensors defined in Sec. \ref{s_thin_film_theory}.
One performs Legendre transformation
\begin{equation}\label{eq_app_leg_transform}
G = F - \sum_{i=1}^6 \epsilon_i \sigma_i
\end{equation}
where $\sum \epsilon_i \sigma_i$ represents the work required to maintain a constant stress.
Then, using the mechanical equation of state
\begin{equation}\label{eq_app_mech_eq}
\sigma_i = -\frac{\partial{F}}{\partial{\epsilon_i}}, i=1..6
\end{equation}
one eliminates stresses $\epsilon_i$ between \eqref{eq_app_leg_transform} and \eqref{eq_app_mech_eq}:
solutions to Eqs. \eqref{eq_app_mech_eq} were expanded in series and only low order terms of the expansion were kept, after that, $\epsilon_i$ were substituted into \eqref{eq_app_leg_transform}.

The transformations of the corresponding high-order electrostrictive coefficients are listed below (\ref{eq_app_m111_transform} - \ref{eq_app_r155_transform}).
\scriptsize
\begin{eqnarray}\label{eq_app_m111_transform}
M_{111}=\frac{\left(c_{11}+c_{12}\right){}^2 m_{111}}{\left(c_{11}-c_{12}\right){}^2 \left(c_{11}+2 c_{12}\right){}^2}-\frac{4
c_{12} \left(c_{11}+c_{12}\right) m_{112}}{\left(c_{11}-c_{12}\right){}^2 \left(c_{11}+2 c_{12}\right){}^2}+ \\ \nonumber
\frac{2 c_{12}^2 m_{122}}{\left(c_{11}-c_{12}\right){}^2
\left(c_{11}+2 c_{12}\right){}^2}+\frac{2 c_{12}^2 m_{123}}{\left(c_{11}-c_{12}\right){}^2 \left(c_{11}+2 c_{12}\right){}^2},
\end{eqnarray}
\begin{eqnarray}\label{eq_app_m112_transform}
M_{112}=-\frac{c_{12} \left(c_{11}+c_{12}\right) m_{111}}{\left(c_{11}-c_{12}\right){}^2 \left(c_{11}+2 c_{12}\right){}^2}+\frac{\left(c_{11}^2+c_{11}
c_{12}+2 c_{12}^2\right) m_{112}}{\left(c_{11}-c_{12}\right){}^2 \left(c_{11}+2 c_{12}\right){}^2}-\\ \nonumber
\frac{c_{11} c_{12} m_{122}}{\left(c_{11}-c_{12}\right){}^2
\left(c_{11}+2 c_{12}\right){}^2}-\frac{c_{11} c_{12} m_{123}}{\left(c_{11}-c_{12}\right){}^2 \left(c_{11}+2 c_{12}\right){}^2},
\end{eqnarray}
\begin{eqnarray}\label{eq_app_m122_transform}
M_{122}=\frac{c_{12}^2 m_{111}}{\left(c_{11}-c_{12}\right){}^2 \left(c_{11}+2 c_{12}\right){}^2}-\frac{2c_{11} c_{12} m_{112}}{\left(c_{11}-c_{12}\right){}^2
\left(c_{11}+2 c_{12}\right){}^2}+ \\ \nonumber
\frac{\left(c_{11}^2+2 c_{11} c_{12}+2 c_{12}^2\right) m_{122}}{\left(c_{11}-c_{12}\right){}^2 \left(c_{11}+2 c_{12}\right){}^2}-\frac{2c_{12}
\left(c_{11}+c_{12}\right) m_{123}}{\left(c_{11}-c_{12}\right){}^2 \left(c_{11}+2 c_{12}\right){}^2},
\end{eqnarray}
\begin{eqnarray}\label{eq_app_m123_transform}
M_{123}=\frac{c_{12}^2 m_{111}}{\left(c_{11}-c_{12}\right){}^2 \left(c_{11}+2 c_{12}\right){}^2}-\frac{2 c_{11} c_{12} m_{112}}{\left(c_{11}-c_{12}\right){}^2
\left(c_{11}+2 c_{12}\right){}^2}-\\ \nonumber
\frac{2 c_{12} \left(c_{11}+c_{12}\right) m_{122}}{\left(c_{11}-c_{12}\right){}^2 \left(c_{11}+2 c_{12}\right){}^2}+\frac{\left(c_{11}^2+2
c_{11} c_{12}+2 c_{12}^2\right) m_{123}}{\left(c_{11}-c_{12}\right){}^2 \left(c_{11}+2 c_{12}\right){}^2},
\end{eqnarray}
\begin{eqnarray}\label{eq_app_m144_transform}
M_{144}=\frac{m_{144}}{c_{44}{}^2},
\end{eqnarray}
\begin{eqnarray}\label{eq_app_m155_transform}
M_{155}=\frac{m_{155}}{c_{44}{}^2},
\end{eqnarray}
\begin{eqnarray}\label{eq_app_m414_transform}
M_{414}=\frac{\left(c_{11}+c_{12}\right) m_{414}}{\left(c_{11}-c_{12}\right) \left(c_{11}+2 c_{12}\right) c_{44}}-\frac{2 c_{12} m_{424}}{\left(c_{11}{}^2+c_{11}
c_{12}-2 c_{12}{}^2\right) c_{44}},
\end{eqnarray}
\begin{eqnarray}\label{eq_app_m424_transform}
M_{424} = -\frac{c_{12} m_{414}}{\left(c_{11}{}^2+c_{11} c_{12}-2 c_{12}{}^2\right) c_{44}}+\frac{c_{11} m_{424}}{\left(c_{11}{}^2+c_{11} c_{12}-2
c_{12}{}^2\right) c_{44}},
\end{eqnarray}
\begin{eqnarray}\label{eq_app_m456_transform}
M_{456}=\frac{m_{456}}{c_{44}{}^2},
\end{eqnarray}
\begin{eqnarray}\label{eq_app_r111_transform}
R_{111}=\frac{\left(c_{11}+c_{12}\right) m_{111} \left(c_{11} q_{11}+c_{12} \left(q_{11}-2 q_{12}\right)\right)}{\left(c_{11}-c_{12}\right){}^2
\left(c_{11}+2 c_{12}\right){}^2}+\frac{2 c_{12} m_{122} \left(c_{12} q_{11}-c_{11} q_{12}\right)}{\left(c_{11}-c_{12}\right){}^2 \left(c_{11}+2
c_{12}\right){}^2}+ \\ \nonumber
\frac{2 c_{12} m_{123} \left(c_{12} q_{11}-c_{11} q_{12}\right)}{\left(c_{11}-c_{12}\right){}^2 \left(c_{11}+2 c_{12}\right){}^2}+\frac{2
m_{112} \left(c_{11}^2 q_{12}+c_{11} c_{12} \left(-2 q_{11}+q_{12}\right)+2 c_{12}^2 \left(-q_{11}+q_{12}\right)\right)}{\left(c_{11}-c_{12}\right){}^2
\left(c_{11}+2 c_{12}\right){}^2}+\\ \nonumber
\frac{\left(c_{11}+c_{12}\right) r_{111}}{\left(c_{11}-c_{12}\right) \left(c_{11}+2 c_{12}\right)}-\frac{2 c_{12}
r_{112}}{c_{11}^2+c_{11} c_{12}-2 c_{12}^2},
\end{eqnarray}
\begin{eqnarray}\label{eq_app_r121_transform}
R_{121}=\frac{m_{122} \left(c_{11}^2 q_{11}+2 c_{12}^2 q_{11}+c_{11} c_{12} \left(q_{11}-4 q_{12}\right)\right)}{2 \left(c_{11}-c_{12}\right){}^2
\left(c_{11}+2 c_{12}\right){}^2}+\frac{c_{11} m_{111} \left(-c_{12} q_{11}+c_{11} q_{12}\right)}{2 \left(c_{11}-c_{12}\right){}^2 \left(c_{11}+2
c_{12}\right){}^2}+\\ \nonumber
\frac{m_{123} \left(-3 c_{11} c_{12} q_{11}-2 c_{12}^2 \left(q_{11}-2 q_{12}\right)+c_{11}^2 q_{12}\right)}{2 \left(c_{11}-c_{12}\right){}^2
\left(c_{11}+2 c_{12}\right){}^2}+\frac{m_{112} \left(2 c_{12}^2 q_{11}+c_{11}^2 \left(q_{11}+2 q_{12}\right)-c_{11} c_{12} \left(q_{11}+4 q_{12}\right)\right)}{2
\left(c_{11}-c_{12}\right){}^2 \left(c_{11}+2 c_{12}\right){}^2}-\\ \nonumber
\frac{c_{12} m_{414} q_{44}}{2 \left(c_{11}^2+c_{11} c_{12}-2 c_{12}^2\right) c_{44}}+\frac{c_{11}
m_{424} q_{44}}{2 c_{11}^2 c_{44}+2 c_{11} c_{12} c_{44}-4 c_{12}^2 c_{44}}+\frac{c_{11} r_{121}}{c_{11}^2+c_{11} c_{12}-2 c_{12}^2}-\frac{c_{12}
r_{123}}{c_{11}^2+c_{11} c_{12}-2 c_{12}^2},
\end{eqnarray}
\begin{eqnarray}\label{eq_app_r221_transform}
R_{112}=\frac{m_{112} \left(c_{11}^2 q_{11}+2 c_{12}^2 q_{11}+c_{11} c_{12} \left(q_{11}-4 q_{12}\right)\right)}{\left(c_{11}-c_{12}\right){}^2
\left(c_{11}+2 c_{12}\right){}^2}-\frac{c_{12} m_{111} \left(c_{11} q_{11}+c_{12} \left(q_{11}-2 q_{12}\right)\right)}{\left(c_{11}-c_{12}\right){}^2
\left(c_{11}+2 c_{12}\right){}^2}+\\ \nonumber
\frac{c_{11} m_{122} \left(-c_{12} q_{11}+c_{11} q_{12}\right)}{\left(c_{11}-c_{12}\right){}^2 \left(c_{11}+2 c_{12}\right){}^2}+\frac{c_{11}
m_{123} \left(-c_{12} q_{11}+c_{11} q_{12}\right)}{\left(c_{11}-c_{12}\right){}^2 \left(c_{11}+2 c_{12}\right){}^2}-\frac{c_{12} r_{111}}{c_{11}^2+c_{11}
c_{12}-2 c_{12}^2}+\frac{c_{11} r_{112}}{c_{11}^2+c_{11} c_{12}-2 c_{12}^2},
\end{eqnarray}
\begin{eqnarray}\label{eq_app_r123_transform}
R_{123}=\frac{m_{123} \left(c_{11}^2 q_{11}+c_{11} c_{12} \left(2 q_{11}-3 q_{12}\right)+2 c_{12}^2 \left(q_{11}-q_{12}\right)\right)}{\left(c_{11}-c_{12}\right){}^2
\left(c_{11}+2 c_{12}\right){}^2}+\frac{c_{12} m_{111} \left(c_{12} q_{11}-c_{11} q_{12}\right)}{\left(c_{11}-c_{12}\right){}^2 \left(c_{11}+2 c_{12}\right){}^2}+ \\ \nonumber
\frac{m_{122}\left(c_{11}^2 q_{12}+c_{11} c_{12} \left(-2 q_{11}+q_{12}\right)+2 c_{12}^2 \left(-q_{11}+q_{12}\right)\right)}{\left(c_{11}-c_{12}\right){}^2 \left(c_{11}+2
c_{12}\right){}^2}+\frac{m_{112} \left(c_{11}^2 q_{12}+2 c_{12}^2 q_{12}-c_{11} c_{12} \left(2 q_{11}+q_{12}\right)\right)}{\left(c_{11}-c_{12}\right){}^2
\left(c_{11}+2 c_{12}\right){}^2}+ \\ \nonumber
\frac{\left(c_{11}+c_{12}\right) m_{414} q_{44}}{2 \left(c_{11}-c_{12}\right) \left(c_{11}+2 c_{12}\right) c_{44}}-\frac{c_{12}
m_{424} q_{44}}{\left(c_{11}^2+c_{11} c_{12}-2 c_{12}^2\right) c_{44}}-\frac{2 c_{12} r_{121}}{c_{11}^2+c_{11} c_{12}-2 c_{12}^2}+\frac{\left(c_{11}+c_{12}\right)
r_{123}}{\left(c_{11}-c_{12}\right) \left(c_{11}+2 c_{12}\right)},
\end{eqnarray}
\begin{eqnarray}\label{eq_app_r144_transform}
R_{144}=\frac{m_{414} \left(c_{11} q_{11}+c_{12} \left(q_{11}-2 q_{12}\right)\right)}{2 \left(c_{11}^2+c_{11} c_{12}-2 c_{12}^2\right) c_{44}}+\frac{-c_{12}
m_{424} q_{11}+c_{11} m_{424} q_{12}}{\left(c_{11}-c_{12}\right) \left(c_{11}+2 c_{12}\right) c_{44}}+\frac{m_{144} q_{44}}{2 c_{44}^2}+\frac{m_{456}
q_{44}}{c_{44}^2}+\frac{r_{144}}{c_{44}},
\end{eqnarray}
\begin{eqnarray}\label{eq_app_r155_transform}
R_{155}=\frac{m_{414}(c_{11} q_{12}-c_{12} q_{11})}{\left(c_{11}-c_{12}\right) \left(c_{11}+2 c_{12}\right) c_{44}}+\frac{m_{424} \left(-2
c_{12} q_{12}+c_{11} \left(q_{11}+q_{12}\right)\right)}{\left(c_{11}-c_{12}\right) \left(c_{11}+2 c_{12}\right) c_{44}}+\frac{m_{155} q_{44}}{c_{44}^2}+\frac{r_{155}}{c_{44}}.
\end{eqnarray}
\normalsize

\section{Bulk - thin film renormalization of $a_{ijk}$ coefficients}\label{app_bulk_film}
To obtain the effective potential of the film
\begin{equation}\label{eq_app_G_film_a}
\widetilde{G} = G+\epsilon_1 \sigma_1+\epsilon_2 \sigma_2+\epsilon_6 \sigma_6,
\end{equation}
one applies mixed mechanical conditions
\begin{equation}\label{eq_app_G_film_b}
\begin{cases}
\frac{\partial G}{\partial \sigma_1}=\epsilon_0,\frac{\partial G}{\partial \sigma_2}=\epsilon_0,\frac{\partial G}{\partial \sigma_6}=0\\
\sigma_3=0,\sigma_4=0,\sigma_5=0.
\end{cases}
\end{equation}
and eliminates $\sigma_1$, $\sigma_2$ and $\sigma_6$ between \eqref{eq_app_G_film_a} and \eqref{eq_app_G_film_b}.
The high-order couplings described by the $M_{ijk}$, $R_{ijk}$, and $N_{ijk}$ tensors renormalize the $a_{ijk}$ coefficients of the $P^6$-terms of $G$. The renormalization reads:
\scriptsize
\begin{eqnarray}\label{eq_app_a111_transform}
a^\text{film}_{111} = a_{111}^\text{bulk} - \frac{M_{111} \left(Q_{11} s_{11}-Q_{12} s_{12}\right){}^2}{2 \left(s_{11}-s_{12}\right){}^2 \left(s_{11}+s_{12}\right){}^2}-\frac{M_{122}
\left(Q_{12}^2 s_{11}^2-2 Q_{11} Q_{12} s_{11} s_{12}+Q_{11}^2 s_{12}^2\right)}{2 \left(s_{11}-s_{12}\right){}^2 \left(s_{11}+s_{12}\right){}^2}- ~~~~~\\ \nonumber
\frac{R_{112}\left(-Q_{12} s_{11}^3+Q_{11} s_{11}^2 s_{12}+Q_{12} s_{11} s_{12}^2-Q_{11} s_{12}^3\right)}{\left(s_{11}-s_{12}\right){}^2 \left(s_{11}+s_{12}\right){}^2}-\frac{R_{111}
\left(-Q_{11} s_{11}^3+Q_{12} s_{11}^2 s_{12}+Q_{11} s_{11} s_{12}^2-Q_{12} s_{12}^3\right)}{\left(s_{11}-s_{12}\right){}^2 \left(s_{11}+s_{12}\right){}^2}- \\ \nonumber
\frac{M_{112}\left(-Q_{11}^2 s_{11} s_{12}-Q_{12}^2 s_{11} s_{12}+Q_{11} Q_{12} \left(s_{11}^2+s_{12}^2\right)\right)}{\left(s_{11}-s_{12}\right){}^2 \left(s_{11}+s_{12}\right){}^2}+ \\ \nonumber
\frac{\left(Q_{11}+Q_{12}\right)}{3 \left(s_{11}-s_{12}\right){}^2
\left(s_{11}+s_{12}\right){}^3} (Q_{11}{}^2 \left(s_{11}{}^2 N_{111}+s_{11} \left(N_{111}-3 N_{112}\right) s_{12}+N_{111}
s_{12}{}^2\right)+ \\ \nonumber
Q_{12}{}^2 \left(s_{11}{}^2 N_{111}+s_{11} \left(N_{111}-3 N_{112}\right) s_{12}+N_{111} s_{12}{}^2\right)- \\ \nonumber
Q_{11} Q_{12} \left(s_{11}{}^2
\left(N_{111}-3 N_{112}\right)+4 s_{11} N_{111} s_{12}+\left(N_{111}-3 N_{112}\right) s_{12}{}^2\right)),
\end{eqnarray}
\begin{eqnarray}\label{eq_app_a333_transform}
a^\text{film}_{333} = a_{111}^\text{bulk} - \frac{M_{122} Q_{12}^2}{\left(s_{11}+s_{12}\right){}^2}-\frac{M_{123} Q_{12}^2}{\left(s_{11}+s_{12}\right){}^2}+\frac{2 Q_{12}
R_{112}}{s_{11}+s_{12}}+ \frac{2 Q_{12}{}^3 \left(N_{111}+3 N_{112}\right)}{3 \left(s_{11}+s_{12}\right){}^3},
\end{eqnarray}
\begin{eqnarray}\label{eq_app_a112_transform}
a^\text{film}_{112} = a_{112}^\text{bulk} - \frac{M_{122} \left(Q_{11}^2 s_{11} \left(s_{11}-2 s_{12}\right)+Q_{12}^2 s_{12}
\left(-2 s_{11}+s_{12}\right)+2 Q_{11} Q_{12} \left(s_{11}^2-s_{11} s_{12}+s_{12}^2\right)\right)}{2 \left(s_{11}-s_{12}\right){}^2 \left(s_{11}+s_{12}\right){}^2}- ~~~~~ \\ \nonumber
\frac{M_{112}\left(Q_{11} Q_{12} \left(s_{11}^2-4 s_{11} s_{12}+s_{12}^2\right)+Q_{11}^2 \left(s_{11}^2-s_{11} s_{12}+s_{12}^2\right)+Q_{12}^2 \left(s_{11}^2-s_{11}
s_{12}+s_{12}^2\right)\right)}{\left(s_{11}-s_{12}\right){}^2 \left(s_{11}+s_{12}\right){}^2}- \\ \nonumber
\frac{M_{155} Q_{44}^2 \left(s_{11}^2-s_{12}^2\right){}^2}{2
\left(s_{11}-s_{12}\right){}^2 \left(s_{11}+s_{12}\right){}^2 s_{44}^2}-\frac{M_{424} \left(Q_{11}+Q_{12}\right) Q_{44}}{\left(s_{11}+s_{12}\right)
s_{44}}+\frac{Q_{44} R_{155} \left(s_{11}^2-s_{12}^2\right){}^2}{\left(s_{11}-s_{12}\right){}^2 \left(s_{11}+s_{12}\right){}^2 s_{44}}- \\ \nonumber
\frac{M_{111}\left(2 Q_{11} Q_{12} s_{11} s_{44}^2(s_{11}-s_{12}) + Q_{12}^2 s_{11}^2 s_{44}^2-2 Q_{11}^2 s_{11} s_{12} s_{44}^2-2 Q_{11} Q_{12} s_{11} s_{12} s_{44}^2-2 Q_{12}^2
s_{11} s_{12} s_{44}^2+Q_{11}^2 s_{12}^2 s_{44}^2\right)}{2 \left(s_{11}-s_{12}\right){}^2 \left(s_{11}+s_{12}\right){}^2
s_{44}^2}- \\ \nonumber
\frac{R_{111} \left(-Q_{12} s_{11}^3 s_{44}^2+Q_{11} s_{11}^2 s_{12} s_{44}^2+Q_{12} s_{11} s_{12}^2 s_{44}^2-Q_{11} s_{12}^3 s_{44}^2\right)}{\left(s_{11}-s_{12}\right){}^2
\left(s_{11}+s_{12}\right){}^2 s_{44}^2}-\\ \nonumber
(R_{121} (-2 Q_{11} s_{11}^3 s_{44}^2-2 Q_{12} s_{11}^3 s_{44}^2+2 Q_{11} s_{11}^2 s_{12} s_{44}^2+2
Q_{12} s_{11}^2 s_{12} s_{44}^2+ 2 Q_{11} s_{11} s_{12}^2 s_{44}^2+2 Q_{12} s_{11} s_{12}^2 s_{44}^2- \\ \nonumber
2 Q_{11} s_{12}^3 s_{44}^2-2 Q_{12} s_{12}^3
s_{44}^2))/(\left(s_{11}-s_{12}\right){}^2 \left(s_{11}+s_{12}\right){}^2 s_{44}^2)-\\ \nonumber
\frac{R_{112} \left(-Q_{11} s_{11}^3 s_{44}^2+Q_{12} s_{11}^2
s_{12} s_{44}^2+Q_{11} s_{11} s_{12}^2 s_{44}^2-Q_{12} s_{12}^3 s_{44}^2\right)}{\left(s_{11}-s_{12}\right){}^2 \left(s_{11}+s_{12}\right){}^2 s_{44}^2} + \\ \nonumber
\frac{\left(Q_{11}+Q_{12}\right)}{\left(s_{11}-s_{12}\right){}^2 \left(s_{11}+s_{12}\right){}^3 s_{44}{}^2}\left(Q_{44}{}^2\right(s_{11}{}^2-s_{12}{}^2)^2N_{155}
\text{+(}Q_{11}{}^2(s_{11}{}^2N_{112}-s_{11}(N_{111}+N_{112})s_{12}+N_{112}s_{12}{}^2\text{)+} \\ \nonumber
Q_{12}{}^2(s_{11}{}^2
N_{112}-s_{11}(N_{111}+N_{112})s_{12}+N_{112}s_{12}{}^2\text{)+}Q_{11} Q_{12}(s_{11}{}^2(N_{111}+N_{112}\text{)- 4}s_{11}N_{112}s_{12}\text{+(}N_{111}+N_{112})s_{12}{}^2\text{))}s_{44}{}^2,
\end{eqnarray}
\begin{eqnarray}\label{eq_app_a113_transform}
a^\text{film}_{113} = a_{112}^\text{bulk} - \frac{M_{111} Q_{12} \left(Q_{11} s_{11}-Q_{12} s_{12}\right)}{\left(s_{11}-s_{12}\right) \left(s_{11}+s_{12}\right){}^2}-\frac{M_{123}
\left(Q_{11} Q_{12} s_{11}^2-Q_{11}^2 s_{11} s_{12}-Q_{12}^2 s_{11} s_{12}+Q_{11} Q_{12} s_{12}^2\right)}{\left(s_{11}-s_{12}\right){}^2 \left(s_{11}+s_{12}\right){}^2}- \\ \nonumber
\frac{M_{112}\left(Q_{11} Q_{12} s_{11}^2+Q_{12}^2 s_{11}^2-2 Q_{11} Q_{12} s_{11} s_{12}-2 Q_{12}^2 s_{11} s_{12}+Q_{11} Q_{12} s_{12}^2+Q_{12}^2 s_{12}^2\right)}{\left(s_{11}-s_{12}\right){}^2
\left(s_{11}+s_{12}\right){}^2}-\\ \nonumber
\frac{R_{123} \left(-2 Q_{12} s_{11}^3+2 Q_{11} s_{11}^2 s_{12}+2 Q_{12} s_{11} s_{12}^2-2 Q_{11} s_{12}^3\right)}{\left(s_{11}-s_{12}\right){}^2
\left(s_{11}+s_{12}\right){}^2}-\\ \nonumber
\frac{R_{121} \left(-2 Q_{11} s_{11}^3+2 Q_{12} s_{11}^2 s_{12}+2 Q_{11} s_{11} s_{12}^2-2 Q_{12} s_{12}^3\right)}{\left(s_{11}-s_{12}\right){}^2
\left(s_{11}+s_{12}\right){}^2}-\\ \nonumber
\frac{R_{111} \left(-Q_{12} s_{11}^3+Q_{12} s_{11}^2 s_{12}+Q_{12} s_{11} s_{12}^2-Q_{12} s_{12}^3\right)}{\left(s_{11}-s_{12}\right){}^2
\left(s_{11}+s_{12}\right){}^2}-\frac{R_{112} \left(-Q_{12} s_{11}^3+Q_{12} s_{11}^2 s_{12}+Q_{12} s_{11} s_{12}^2-Q_{12} s_{12}^3\right)}{\left(s_{11}-s_{12}\right){}^2
\left(s_{11}+s_{12}\right){}^2}-\\ \nonumber
\frac{M_{122} \left(2 Q_{11} Q_{12} s_{12} \left(-3 s_{11}+s_{12}\right)+Q_{11}^2 \left(s_{11}^2+s_{12}^2\right)+Q_{12}^2
\left(3 s_{11}^2-2 s_{11} s_{12}+s_{12}^2\right)\right)}{2 \left(s_{11}-s_{12}\right){}^2 \left(s_{11}+s_{12}\right){}^2} + \\ \nonumber
\frac{Q_{12}}{\left(s_{11}-s_{12}\right){}^2\left(s_{11}+s_{12}\right){}^3}\text{(4 }Q_{11}Q_{12}(s_{11}{}^2N_{112}-s_{11}(N_{111}+N_{112})s_{12}+N_{112}s_{12}{}^2\text{)+}Q_{11}{}^2(s_{11}{}^2(N_{111}+N_{112} )- \\ \nonumber
4s_{11}N_{112}s_{12}\text{+(}N_{111}+N_{112})s_{12}{}^2\text{)+}Q_{12}{}^2(s_{11}{}^2(N_{111}+N_{112} - 4s_{11}N_{112}s_{12}\text{+(}N_{111}+N_{112})s_{12}{}^2)),
\end{eqnarray}
\begin{eqnarray}\label{eq_app_a133_transform}
a^\text{film}_{133} = a_{112}^\text{bulk} - \frac{M_{111} Q_{12}^2}{2 \left(s_{11}+s_{12}\right){}^2}-\frac{M_{112} Q_{12}^2}{\left(s_{11}+s_{12}\right){}^2}-\frac{M_{123} Q_{12} \left(Q_{11}+Q_{12}\right)}{\left(s_{11}+s_{12}\right){}^2}-\frac{M_{122}
Q_{12} \left(2 Q_{11}+3 Q_{12}\right)}{2 \left(s_{11}+s_{12}\right){}^2}+\\ \nonumber
\frac{R_{112} \left(Q_{11} s_{11}+Q_{12} s_{11}+Q_{11} s_{12}+Q_{12} s_{12}\right)}{\left(s_{11}+s_{12}\right){}^2}+\frac{R_{121}
\left(2 Q_{12} s_{11}+2 Q_{12} s_{12}\right)}{\left(s_{11}+s_{12}\right){}^2}+\frac{R_{123} \left(2 Q_{12} s_{11}+2 Q_{12} s_{12}\right)}{\left(s_{11}+s_{12}\right){}^2}+ \\ \nonumber
\frac{Q_{12}{}^2 \left(Q_{11}+Q_{12}\right) \left(N_{111}+3 N_{112}\right)}{\left(s_{11}+s_{12}\right){}^3},
\end{eqnarray}
\begin{eqnarray}\label{eq_app_a123_transform}
a^\text{film}_{123} = a_{123}^\text{bulk} - \frac{2 M_{112} Q_{12} \left(Q_{11}+Q_{12}\right)}{\left(s_{11}+s_{12}\right){}^2}+\frac{2 M_{111} Q_{12} \left(-s_{11}+s_{12}\right)
\left(Q_{12} s_{11}-Q_{11} s_{12}\right)}{\left(s_{11}-s_{12}\right){}^2 \left(s_{11}+s_{12}\right){}^2}-\\ \nonumber
\frac{M_{123} \left(-4 Q_{11} Q_{12} s_{11}
s_{12}+Q_{11}^2 \left(s_{11}^2+s_{12}^2\right)+Q_{12}^2 \left(s_{11}^2+s_{12}^2\right)\right)}{\left(s_{11}-s_{12}\right){}^2 \left(s_{11}+s_{12}\right){}^2}+ \\ \nonumber
\frac{2M_{122} \left(Q_{11}^2 s_{11} s_{12}+Q_{12}^2 \left(2 s_{11}-s_{12}\right) s_{12}-Q_{11} Q_{12} \left(2 s_{11}^2-s_{11} s_{12}+s_{12}^2\right)\right)}{\left(s_{11}-s_{12}\right){}^2
\left(s_{11}+s_{12}\right){}^2}-\frac{M_{144} Q_{44}^2 \left(s_{11}^2-s_{12}^2\right){}^2}{2 \left(s_{11}-s_{12}\right){}^2 \left(s_{11}+s_{12}\right){}^2
s_{44}^2}-\\ \nonumber
\frac{2 M_{424} Q_{12} Q_{44}}{\left(s_{11}+s_{12}\right) s_{44}}+\frac{2 Q_{44} R_{144} \left(s_{11}^2-s_{12}^2\right){}^2}{\left(s_{11}-s_{12}\right){}^2
\left(s_{11}+s_{12}\right){}^2 s_{44}}-\\ \nonumber
\frac{R_{123} \left(4 Q_{11} s_{11} \left(s_{11}+s_{12}\right) \left(-s_{11}+s_{12}\right) s_{44}^2+4 Q_{12}
\left(-s_{11}-s_{12}\right) s_{12} \left(-s_{11}+s_{12}\right) s_{44}^2\right)}{\left(s_{11}-s_{12}\right){}^2 \left(s_{11}+s_{12}\right){}^2 s_{44}^2}-\\ \nonumber
(R_{121}(8 Q_{12} s_{11} \left(s_{11}+s_{12}\right) \left(-s_{11}+s_{12}\right) s_{44}^2+4 Q_{11} \left(-s_{11}-s_{12}\right) s_{12} \left(-s_{11}+s_{12}\right)
s_{44}^2+ \\ \nonumber
4 Q_{12} \left(-s_{11}-s_{12}\right) s_{12} \left(-s_{11}+s_{12}\right) s_{44}^2))/(\left(s_{11}-s_{12}\right){}^2 \left(s_{11}+s_{12}\right){}^2 s_{44}^2)+ \\ \nonumber
\frac{2 Q_{12} N_{111} \left(2 Q_{11} Q_{12} s_{11}{}^2 s_{44}{}^2-2 Q_{11}{}^2 s_{11} s_{12} s_{44}{}^2-
2 Q_{12}{}^2 s_{11} s_{12} s_{44}{}^2+2
Q_{11} Q_{12} s_{12}{}^2 s_{44}{}^2\right)}{\left(s_{11}-s_{12}\right){}^2 \left(s_{11}+s_{12}\right){}^3 s_{44}{}^2}+ \\ \nonumber
\frac{2 Q_{12} Q_{44}{}^2 \left(s_{11}{}^2-s_{12}{}^2\right){}^2 N_{155}}{\left(s_{11}-s_{12}\right){}^2 \left(s_{11}+s_{12}\right){}^3
s_{44}{}^2}+ \\ \nonumber
(2 Q_{12} N_{112} (2
Q_{11}{}^2 s_{11}{}^2 s_{44}{}^2+2 Q_{11} Q_{12} s_{11}{}^2 s_{44}{}^2+2 Q_{12}{}^2 s_{11}{}^2 s_{44}{}^2-2 Q_{11}{}^2 s_{11} s_{12} s_{44}{}^2- \\ \nonumber
8Q_{11} Q_{12} s_{11} s_{12} s_{44}{}^2-2 Q_{12}{}^2 s_{11} s_{12} s_{44}{}^2+2 Q_{11}{}^2 s_{12}{}^2 s_{44}{}^2+2 Q_{11} Q_{12} s_{12}{}^2 s_{44}{}^2+2
Q_{12}{}^2 s_{12}{}^2 s_{44}{}^2))/ \\ \nonumber
(\left(s_{11}-s_{12}\right){}^2 \left(s_{11}+s_{12}\right){}^3 s_{44}{}^2)
\end{eqnarray}
\normalsize
Note that in \eqref{eq_app_G_film_b} solutions were expanded in series and only low order terms of the expansion were kept, then $\sigma_1$, $\sigma_2$ and $\sigma_6$ were substituted into \eqref{eq_app_G_film_a}, therefore expressions (\ref{eq_app_a111_transform} - \ref{eq_app_a123_transform}) result from the expansion in series in \eqref{eq_app_G_film_b}.

\end{document}